\documentstyle[epsfig,12pt]{article}
\begin{document}
\begin{flushright}
{\bf  {\em DFUB 2000/01 }}
\end{flushright}

\vspace{1cm}

\begin{center}
{\bf \Large {\em Energy  Losses of Q-balls.}}\\
\end{center}

\begin{center}
{{\em D. Bakari$^{a,b}$, H. Dekhissi$^{a,b}$,
J.Derkaoui$^{a,b}$, 
G. Giacomelli$^{a}$,\\ 
G. Mandrioli$^{a}$, 
M. Ouchrif~$^{a,b}$, L. Patrizii$^{a}$, V. Popa$^{a,c}$.}\\
\vspace{0.5cm}
$^{a}$Dipartimento di Fisica dell' Universit\`a di Bologna, INFN sezione di 
Bologna, Italy.\\
$^{b}$ Facult\'e des sciences (LPTP), Universit\'e Mohamed $1^{er}$, Oujda,
Morocco.\\
$^{c}$ Institute for Space Sciences, Bucharest, Romania.}
\end{center}

\vspace{3cm}
\begin{center}
{\bf \em Abstract}
\end{center}

After a short review of Q-balls properties, in this paper 
we discuss their interaction with 
matter, and their energy losses in the earth, 
for a large range of velocities. These calculations are used to compute
the fractional angular acceptance of a detector at Gran Sasso Laboratory. 
Furthermore 
 we computed the light yield in liquid
scintillators, the ionization in streamer tubes and the Restricted Energy Loss
 in the CR39 nuclear track detector.

\vspace{2cm}
\begin{center}
{\em Submitted to Astroparticle Physics}
\end{center}

\newpage
\vspace{7mm}

\section{Introduction}
Dark Matter (DM) is one of the intriguing problems in particle physics 
and cosmology. Several types of stable (or quasi stable) particles have been
proposed as condidates for the cold DM. Among these condidates one has the 
nuclearites (strange 
quark matter), which are agglomerates of quarks {\it u,d} and {\it s} 
\cite{Witten84}. 
Another example 
of cold DM candidate, coming from theories beyond the standard Model of particle
physics, is the lightest supersymmetric particle (LSP) \cite{TDLEE92}.

In theories where scalar fields
carry a conserved global quantum number,  
$Q$, there may exist non-topological solitons which are stabilized by global  
charge conservation. They 
act like homogenous balls of matter, with $Q$ playing the role of 
the quantum number; Coleman called this type of matter {\em Q-balls}  
\cite{Coleman85}.

The conditions for the existence of absolutely stable Q-balls are 
satisfied in supersymmetric theories with low energy supersymmetry 
breaking\footnote{According to ref. \cite{Kusenko98A}, 
abelian non-topological solitons with baryon and/or lepton 
quantum numbers naturally appear in the spectrum of the Minimal 
Supersymmetric Standard Model.} 
\cite{Kusenko97}. The role of conserved quantum number 
is played by the baryon number  
(The same reasoning applies to the lepton number  for the sleptonic  Q-balls).
Q-balls can be considered like 
coherent states  of squarks, sleptons and Higgs fields.
Under certain assumptions about the internal self interaction of
these particles
and fields the Q-balls could be absolutely stable \cite{Kusenko98A}.

In this note we summarize the main physical and astrophysical properties
of Q-balls,
their interactions with matter, their energy losses in detectors,
and the possibility of traversing the earth to reach any 
detector at the underground Gran Sasso Laboratory (Italy), for example 
MACRO  \cite{Ouchrif98}. We neglect the possibility of : 
$(i)$ electromagnetic radiation emitted by Q-balls of 
high $\beta$, $(ii)$ strong interactions of Q-balls  
in the upper atmosphere capable of destroying the Q-ball, and $(iii)$ the 
possibility that neutral Q-balls (SENS) may become charged (SECS) 
and viceversa \cite{TDLEE92}.

\section{Properties of Q-balls} 

Q-balls could have  
been produced in the Early Universe, and may  contribute to the cold DM. 
Several mechanisms could have lead to the formation of Q-balls in the 
Early Universe. They may have originated 
in the course of a phase transition, which is sometimes  called  
``solitogenesis'', or they could  have  
been produced via fusion processes, 
 reminiscent of 
the Big Bang 
nucleosynthesis,  which have been called 
`` solitosynthesis'';  small Q-balls could be 
pair-produced in very high energy collisions (at high 
temperatures) \cite{Kusenko97}.

The astrophysical 
consequenses of Q-balls in many ways resemble those of  
strange quark matter, nuclearites.

One of the peculiarities of Q-balls is that their 
mass grows as $Q^{3/4}$, while for nuclearites the mass grows linearly with 
the  baryon number \cite{Witten84}.

In the squark bag there is an almost uniform potential $U(\phi)$, which may be 
taken as $U(\phi) \sim M_{S}^4=$constant in SUSY theories with low energy 
supersymmetry breaking \cite{Kusenko98B}. For 
large scalar $\phi$, the mass $M_{Q}$ and radius $R_{Q}$ of Q-balls 
with baryon number $Q$ are given by \cite{Kusenko98B}:

$$
M_{Q}=\frac{4 \pi \sqrt{2}}{3}M_{S}~Q^{3/4} 
\simeq 5924  \left(\frac{M_{S}}{1~TeV}\right)~Q^{3/4}~~~~~(GeV)\\
$$
\begin{equation}
\simeq 1. 2 \times 10^{-20} \left(\frac{M_{S}}{1~TeV}\right)~Q^{3/4}~~~~~(g)
\label{eq:1}
\end{equation}

\begin{equation}
R_{Q}=\frac{1}{\sqrt{2}}M_{S}^{-1}~Q^{1/4} 
\simeq 1.4 \times 10^{-17} \left(\frac{1~TeV}{M_{S}}\right)~Q^{1/4}~~~~~(cm)
\label{eq:2}
\end{equation}
The parameter $M_{S}$ is the energy scale of the SUSY breaking symmetry. 
The first parts of formulae (1), (2) are in $\hbar=c=1$ units; 
the last  parts are in cgs units. 
A stability condition of  
the stable Q-ball mass $M_Q$  
is related to the proton mass $M_{p}$ by 

\begin{equation}
M_{Q} < Q~M_{p} \label{eq:3}
\end{equation}
From Eq. \ref{eq:1} and Eq. \ref{eq:3} one has the stability 
constraint   
\begin{equation}
Q > \frac{M_{Q}}{M_{p}} =1.6 \times 10^{15}\left (\frac{M_{S}}{1~TeV}\right)^4 \label{eq:4}
\end{equation}

\begin{figure}
\begin{center}
        \mbox{ \epsfysize=15cm
            \epsfbox{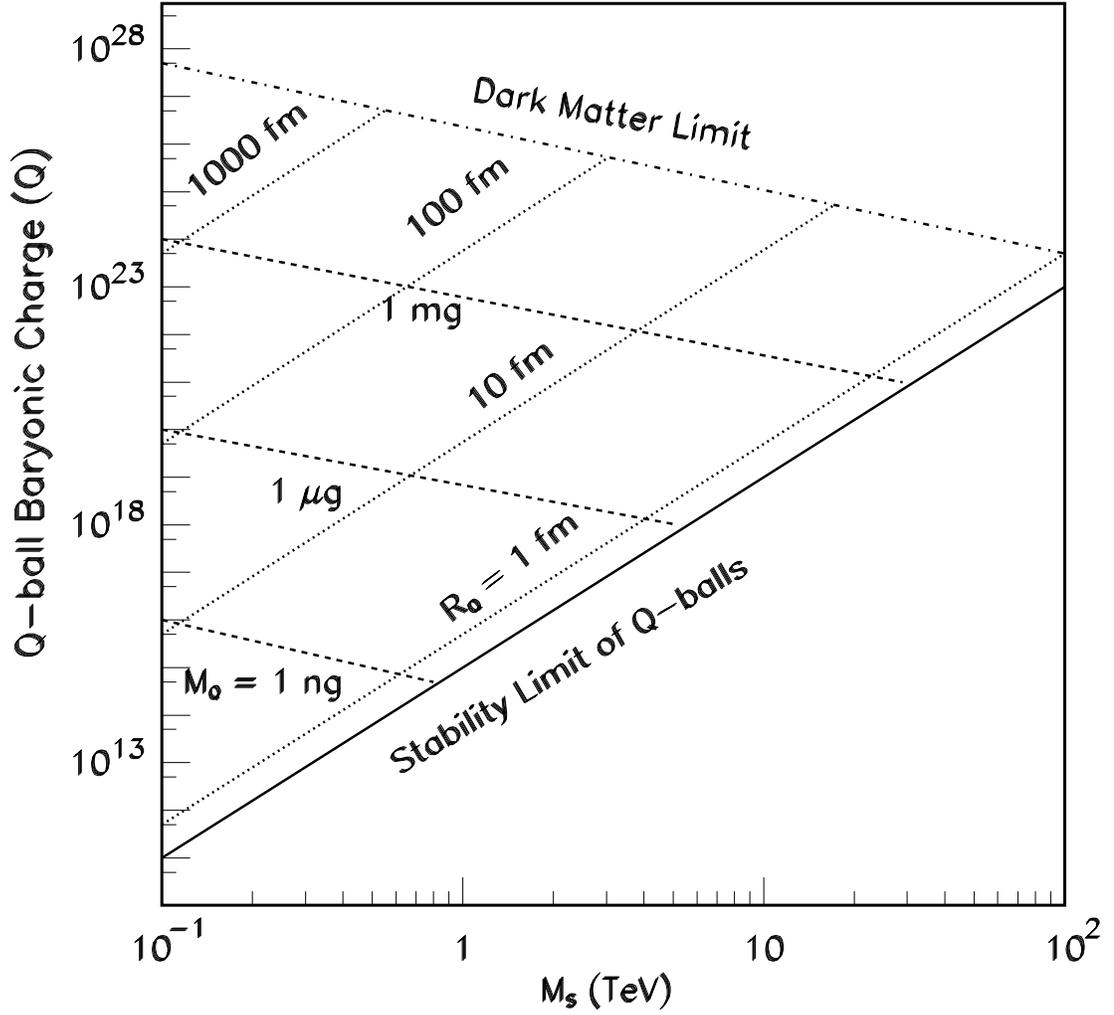}}
\vspace{-1cm}
\caption{Q-ball number versus the supersymmetry energy scale 
$M_S$ for Q-balls. The allowed 
region is delimited from below 
by the Q-ball stability limit (Eq. \ref{eq:4}); assuming that they are part of 
the galactic DM and have $\beta \sim 10^{-3}$, they are limited from above by 
the dashed-dotted  line. 
The dashed lines are $Q=Q(M_{s})$ for different 
Q-balls masses $M_{Q}$; the dotted lines are $Q=Q(M_{s})$ for 
different Q-balls radii $R_{Q}$.}
\label{fig:1}
\end{center}
\end{figure}

In Fig. \ref{fig:1} the allowed region for stable Q-balls is indicated as a shaded 
region in the plane (Q, $M_{S}$). The dashed lines indicate the Q-ball number
as function 
of $M_{S}$ assuming different Q-ball masses $M_{Q}$; the dotted lines represent
Q($M_{S}$) for different Q-balls radii (R).

Relic Q-balls are expected to concentrate in galactic 
halos and to move at the typical galactic velocity 
$v = \beta c \simeq 10^{-3}c$. 
Assuming that Q-balls constitute 
the cold galactic dark matter with $\rho_{DM} \sim 0.3~GeV/cm^{3}$, 
their number density is 
\cite{Kusenko98B}
\begin{equation}
N_{Q} \sim \frac{\rho_{DM}}{M_Q} \\
\sim \rho_{DM}~\left (\frac{3}{4 \pi \sqrt{2}}\right) 
Q^{- 3/4} M_{S}^{-1} \label{eq:5}
\end{equation}
\begin{displaymath}
\sim 5 \times 10^{-5}~Q^{-3/4}
\left(\frac{1~TeV}{M_{S}}\right)~~~~~~cm^{-3} 
\end{displaymath}

The corresponding flux is \cite{Kusenko98B}
$$
\phi \simeq \frac{1}{4~\pi} N_{Q}~v = \frac{\rho_{DM}~v}{4~ \pi M_Q}
$$
\begin{equation}
\simeq \frac{1}{4\pi}~ \rho_{DM}~ 
\left(\frac{4 \pi ~\sqrt{2}}{3} \right)^{-1}Q^{-3/4} M_{S}^{-1}~v \label{eq:6a}
\end{equation}
\begin{displaymath}
\simeq 10^{2}~Q^{-3/4} \left (\frac{M_{S}}{1~TeV} \right )^{-1}v~~~~~~~~~
(cm^{-2}s^{-1}sr^{-1}) ~~~~~~~
\end{displaymath}

In Fig. 2 we present the
dependence of the Q-ball radius $R_Q$ from its mass $M_Q$ for two
values of the $M_S$ parameter. Q-balls with $M_Q \leq 10^8$ GeV  are
unstable; Q-balls with $M_Q \geq 10^{25}$ GeV  should be very rare.

\begin{figure}
\begin{center}
     \mbox{ \epsfysize=16cm
   \epsffile{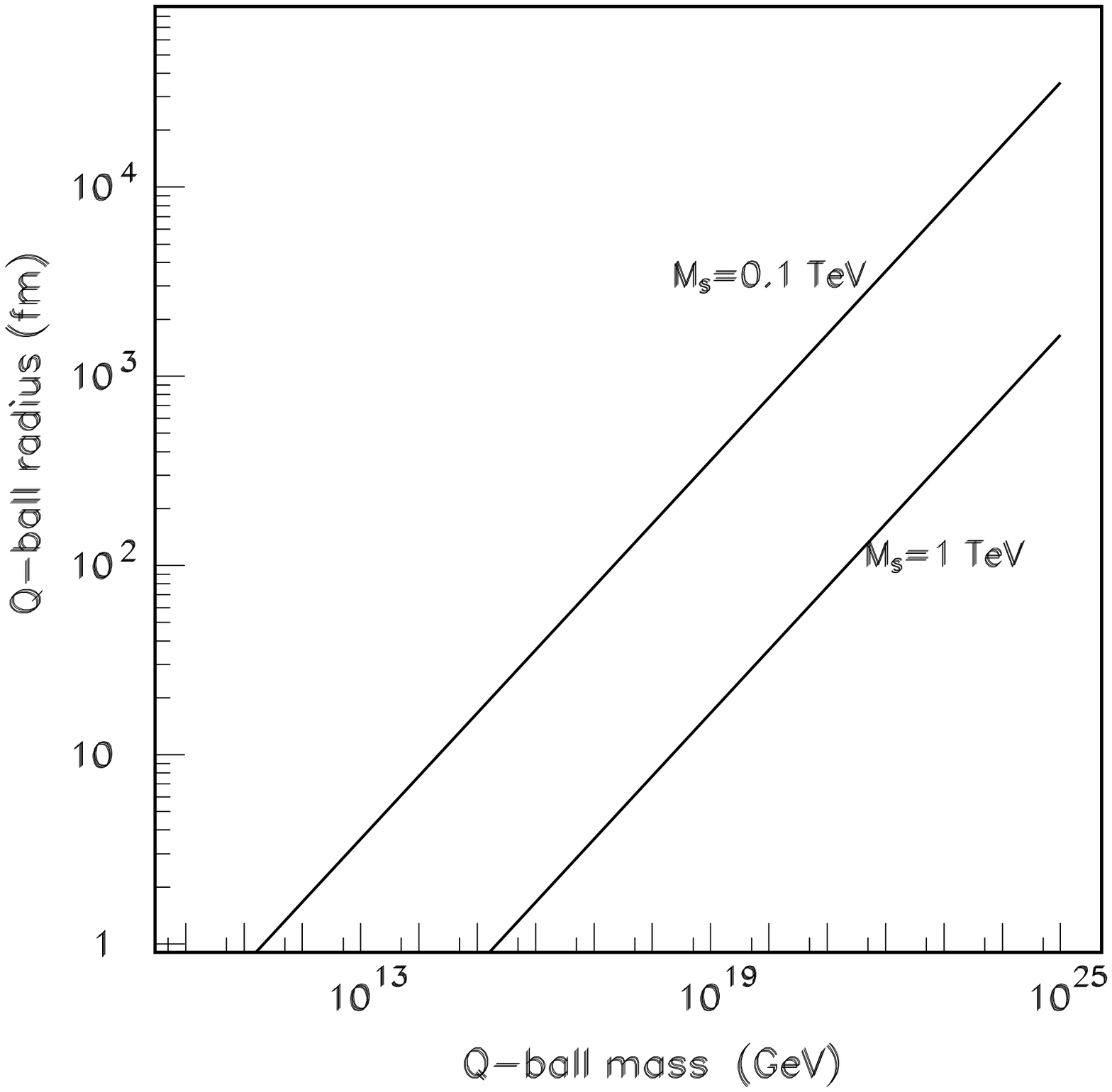}}
\vspace{-1cm}
\caption{Dependence of the Q-ball radius $R_Q$ from its mass $M_Q$ for two
values of the $M_S$ parameter. Q-balls with $M_Q \leq 10^8$ GeV/c$^2$ are
unstable; Q-balls with $M_Q \geq 10^{25}$ GeV/c$^2$ should be very rare.}
\label{fig:2}
\end{center}
\end{figure}

In Fig. 3 we present the 
Q-ball flux versus Q-ball mass assuming that the density of the DM  is 
$\rho =0.3~GeV/cm^3$, that all of it is in the form of Q-balls and that 
$\beta \sim 10^{-3}$. If Q-balls are part of the DM, their flux 
cannot be higher than that of the top dotted-dashed line in Fig.~3.

When traversing normal matter, Q-balls could form 
some sort of bound states with
quarks (which remain in their outer layers), acquiring a positive electric
charge. Electrons around the Q-ball can be captured through the reaction
$ue \rightarrow d\nu_{e}$ or could leave the Q-ball charged if the rate of 
this reaction is small, otherwise they would neutralise it \cite{Kusenko98B}.
\par Q-balls can be classified in two classes:
{\bf SECS} (Supersymmetric Electrically Charged Solitons) and {\bf SENS}  
(Supersymmetric Electrically Neutral Solitons). The interactions 
of Q-balls with matter and their detection differ drastically for SECS 
or SENS \cite{TDLEE92}.

\begin{figure}
\begin{center}
        \mbox{ \epsfysize=16cm
        \epsfbox{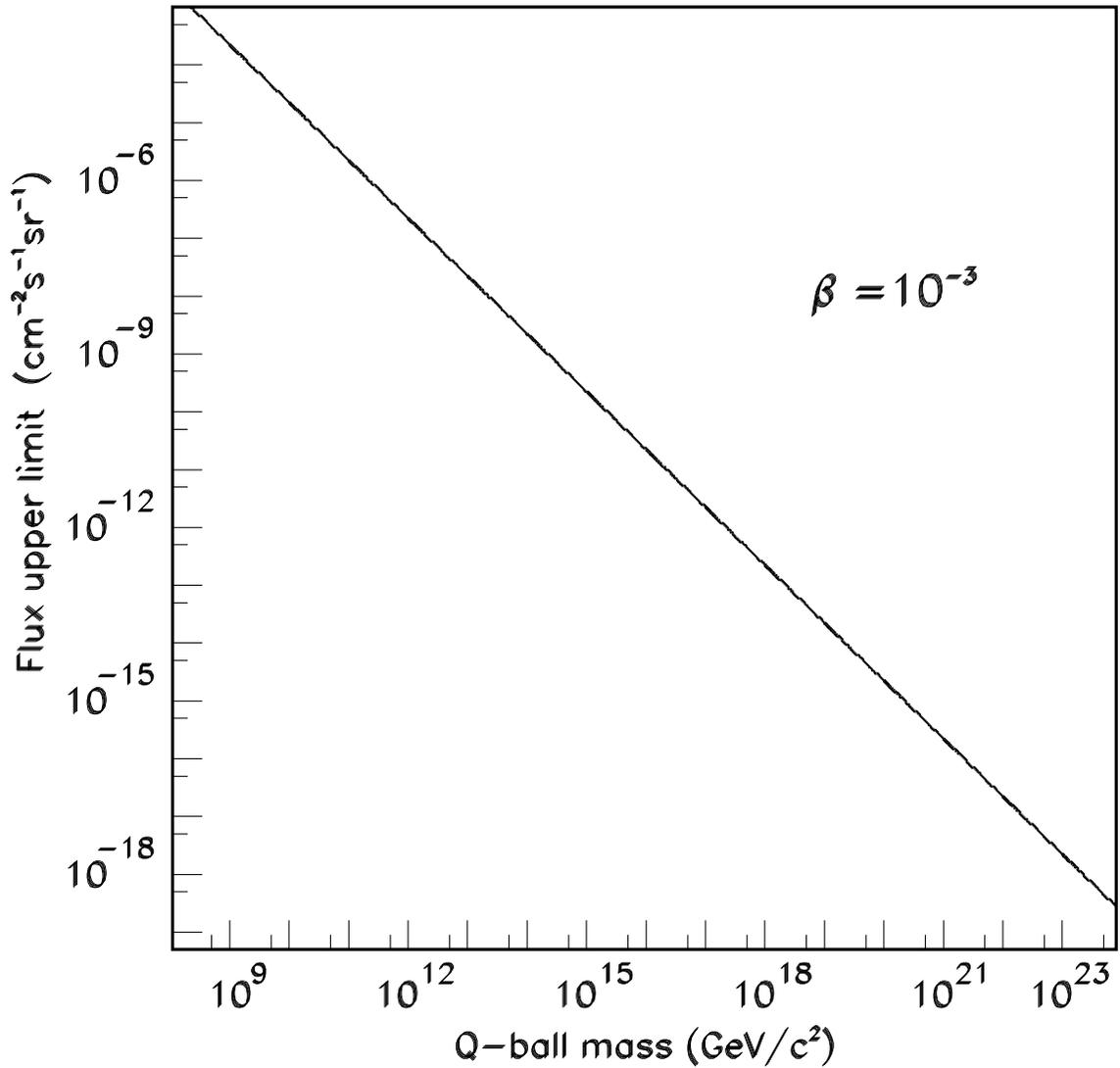}}
\vspace{-1cm}
\caption{Q-ball flux in the cosmic radiation
versus Q-ball mass assuming that the density of DM  is
$\rho =0.3~GeV/cm^3$, that it is all in the form of Q-balls and that
$\beta \sim 10^{-3}$.
Clearly if Q-balls are $0.1\%$ of the total DM, the Q-ball flux is
smaller by a factor of $10^{3}$.}
\label{fig:3}
\end{center}
\end{figure}

\section{Interaction with matter}
Q-balls could have velocities typical of objects bound to the galaxy,
$10^{-4}<\beta<10^{-2}$. Thus we do not expect relativistic Q-balls. For 
reasons of completeness we estimate the energy losses even for relativistic
Q-balls, which we indicate as dotted lines in Figs. 3, 4. 

\subsection{Interactions with matter of Q-balls type SECS}
SECS are Q-balls with a net positive electric charge which tends to be mainly 
in the outer layer. The 
charge of SECS originates from the unequal rates of absorption in the 
condensate. 
The positive  electric charge could be of one unit up to several tens. 
This 
positive electric charge may be
neutralized by a surrounding cloud of electrons.  

For small size Q-balls the positive charge 
interacts with matter (electrons and nuclei) via elastic 
or quasi elastic collisions. 
The cross section is similar to the Bohr cross section of hydrogenoid 
atoms \cite{Kusenko98A}:
\begin{equation}
\sigma = \pi a_{0}^{2}\sim 10^{-16}cm^2 \label{eq:7}
\end{equation}
where $a_{0}$ is the Bohr radius. The formula is valid 
for  $R_{Q} \leq a_{0}$, which happens for $Q \leq 2.7 \times \left(
\frac{M_S}{1~TeV}\right)^4$.
For $Q \geq 2.7 \times \left(\frac{M_S}{1~TeV}\right)^4$ the Q-ball radius 
is larger than $a_{0}$, and it increases with the Q-ball mass. Thus electrons 
could find themselves inside the Q-ball, for large Q-ball masses. One could 
assume that in
such conditions the electronic capture is favoured and thus the SECS tend to 
become SENS.\\

The main energy losses  \cite{Ouchrif98} of SECS passing through 
matter with velocities in the 
range $10^{-4}<\beta<10^{-2}$ are  due to the interaction of the SECS 
positive charge: 
$(i)$ with the nuclei (nuclear contribution),
and $(ii)$ with the electrons of the traversed 
medium  (electronic contribution). 
The total energy loss is the sum of the two 
contributions.

SECS could cause the catalysis of proton decay, but only in the case for   

\begin{equation}
R_{Q} \geq \frac{2Z_{Q}e^2}{m_{e} v^2} \label{eq:8}
\end{equation}
where $v$ and $Z_{Q}$ are respectively 
the velocity and the positive electric charge of the SECS,  
$m_{e}$ is the electron mass [9] (see Fig. \ref{fig:2}). 
For $\beta = 10^{-3}, Z_{Q} = 1$
this corresponds to radii larger than $10^{3}$ fm and masses larger than 1 mg, 
which are very large values. The probability of catalysis by large mass SECS 
is reduced by the presence of their electric charge.

\begin{figure}
\begin{center}
     \mbox{ \epsfysize=15cm
   \epsfbox{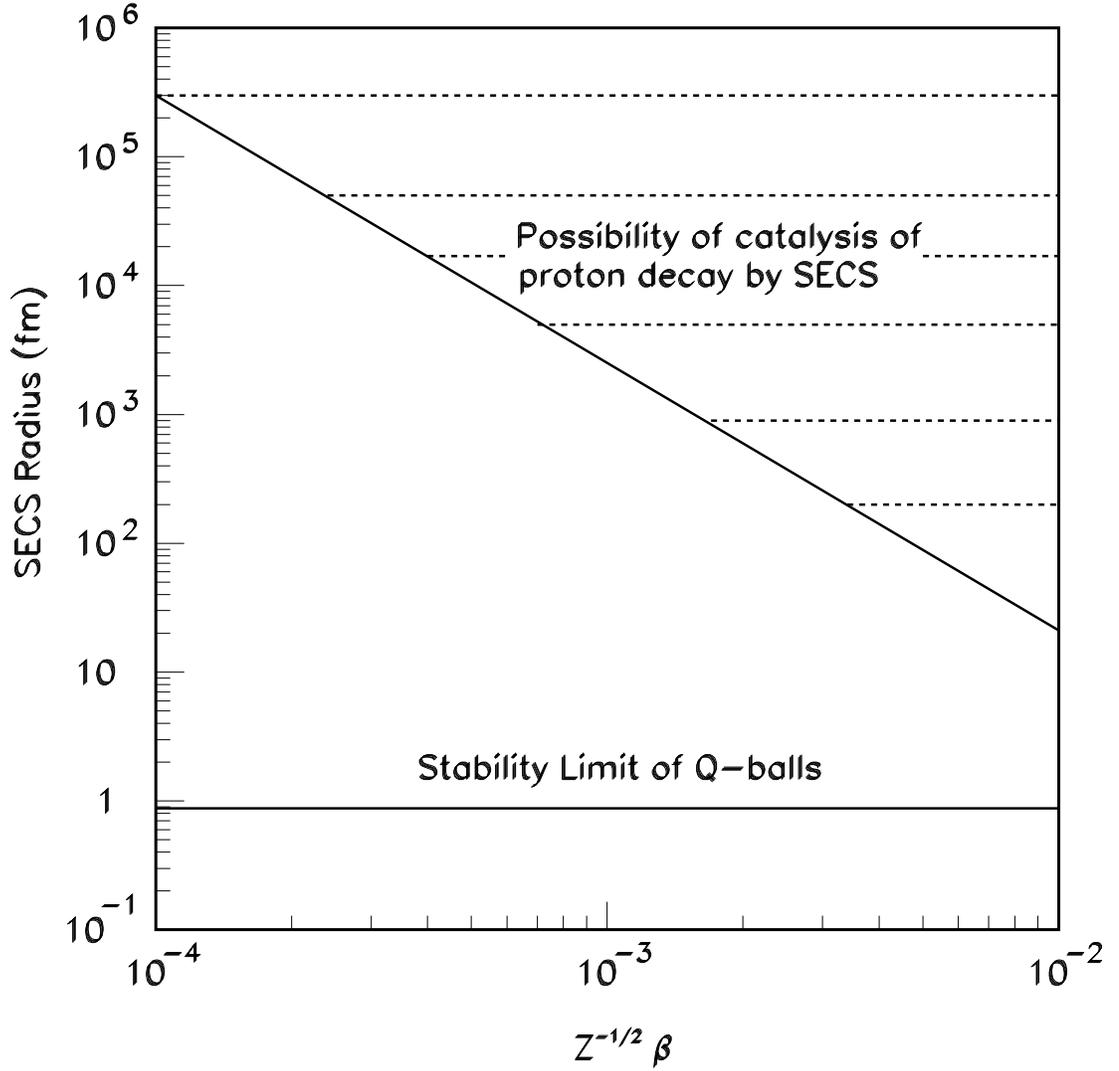}}
\vspace{-1cm}
\caption{For SECS: Q-ball radius $R_{Q}$ versus $Z^{-1/2} \beta$, 
where $Z_{Q}$ 
is the positive 
net charge of the Q-ball core, $\beta = v/c$. 
 The lower line 
indicates the stability limit of Q-balls corresponding approximately to 
a Q-ball mass of $\sim 10^{8}~GeV$.  
The upper solid line gives the minimal Q-ball radii of SECS that could 
catalyse proton decay, Eq. \ref{eq:8}.}
\label{fig:4}
\end{center}
\end{figure}

{\bf Electronic losses of SECS:} The electronic contribution to the energy 
loss of SECS may be computed
with the following formula \cite{Ouchrif98}
\begin{equation}
\frac{dE}{dx} = \frac{8 \pi a_{0} e^{2}  \beta }{\alpha} \frac{Z_{Q}^{7/6} N_{e}}{
(Z_{Q}^{2/3} + Z^{2/3})^{3/2}}~~~~~~~~~for~~Z_{Q} \geq 1 \label{eq:9}
\end{equation}
where $\alpha$ is the fine structure constant, $\beta = v/c$,  
$Z_{Q}$ is the positive charge of SECS, $Z$ is the atomic number 
of the medium and 
$N_e$ is the density of electrons in the medium. Electronic losses dominate 
for $\beta > 10^{-4}$ (see Figs \ref{fig:3} and \ref{fig:4} for the 
case of the dE/dx in the Earth).

{\bf Nuclear losses of SECS:} The nuclear contribution to the energy loss 
of SECS is due to the interaction of the SECS positive charge  
with the nuclei of the medium and it is given by \cite{Ouchrif98}
\begin{equation}
\frac{dE}{dx} = \frac{\pi a^{2} \gamma N E }{\epsilon} S_{n}(\epsilon) 
\label{eq:10}
\end{equation}
where
\begin{equation}
S_{n}(\epsilon) \simeq \frac{0.56 ~ln(1.2\epsilon)}{1.2\epsilon - 
(1.2\epsilon)^{-0.63}}~,~~~\epsilon = \frac{a M E}{Z_{Q}Z e^{2}M_{Q}}
\label{eq:11}
\end{equation}
and
\begin{equation}
a= \frac{0.885~ a_{0}}{(\sqrt{Z_{Q}} + \sqrt{Z})^{2/3}}~,~~~~~~~~~~~~
\gamma= \frac{4 M}{M_{Q}} \label{eq:12}
\end{equation}
$M_{Q}$ is the mass of the incident Q-ball; $M$ is the mass of the 
target nucleus; $Z_{Q}e$ and $Ze$ are their electric charges; $a$ is the 
screening radius and 
$a_{0}$ is the 
Bohr radius; 
we assume 
that $M_{Q} >> M$. Nuclear losses dominates for $\beta \leq 10^{-4}$ (see 
Figs. \ref{fig:3} and \ref{fig:4}).

{\bf The energy losses of SECS in the Earth mantle and Earth core:}
The energy losses of SECS in the Earth mantle and Earth core have been 
computed for different $\beta$-regions and for different charges of the Q-ball 
core, employing the same general procedures used for computing the 
energy losses in the Earth of magnetic monopoles and of nuclearites 
\cite{Ouchrif98}. 
The results 
of the calculations are presented in 
Figs. \ref{fig:3} and \ref{fig:4}; the dashed lines indicate interpolations.

\begin{figure}
\begin{center}
     \mbox{ \epsfysize=15cm
   \epsfbox{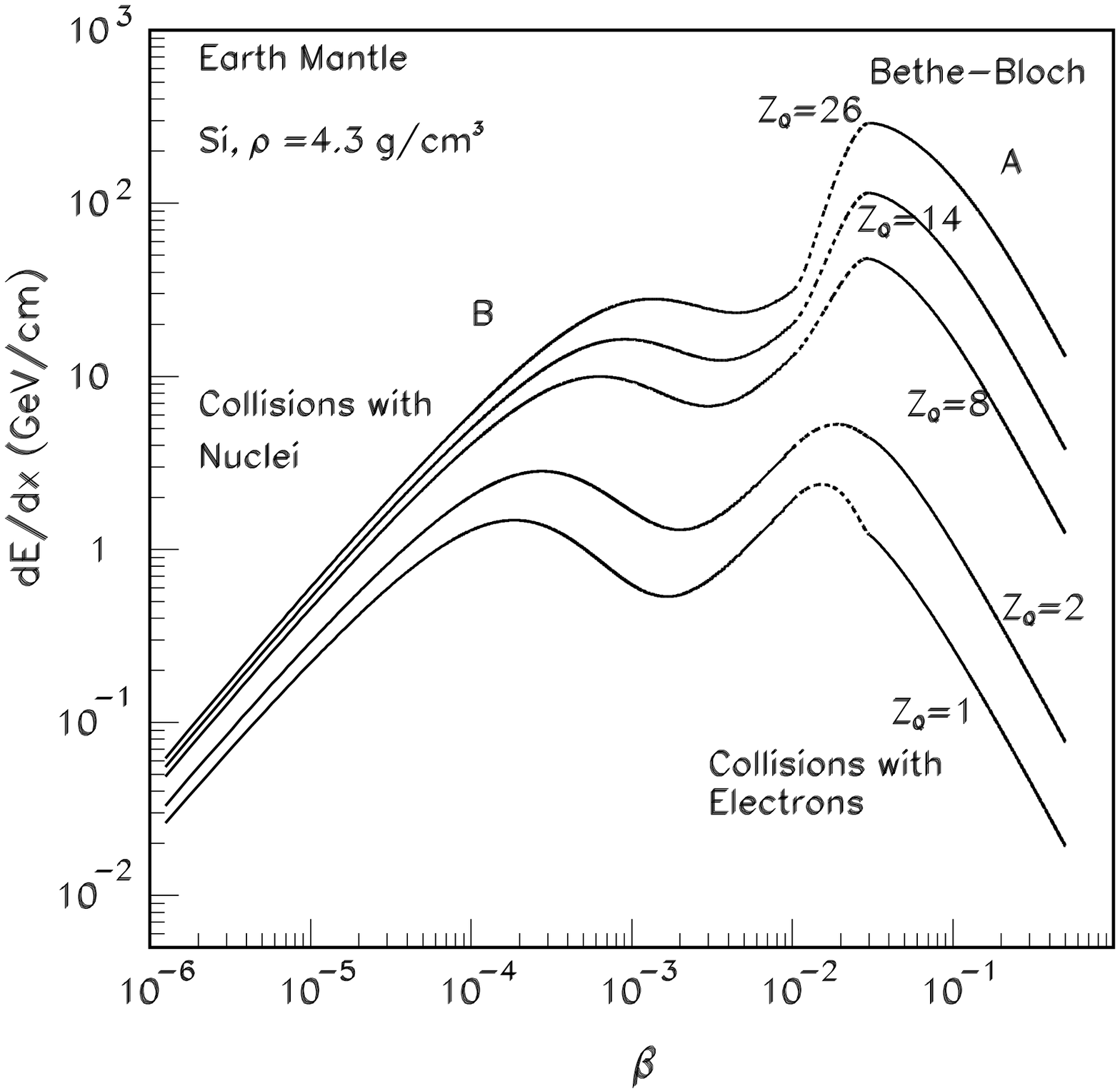}}
\vspace{-1cm}
\caption{Energy losses of SECS versus $\beta$ in the Earth mantle; 
$Z_{Q}$ is the electric charge of the Q-ball. The possibility of the emission 
of electromagnetic radiation at high $\beta \simeq 1$ was not considered.}
\label{fig:5}
\end{center}
\end{figure}

Considering the energy losses discussed above, we have computed 
for a specific velocity 
($v = 250~km/s$)
the angular acceptance ($\Omega/4 \pi$) of the MACRO detector 
(located at an average  depth of 3700 m.w.e.)  
for SECS (with $Z_{Q}=1$), 
see Fig. \ref{fig:5}. Notice that for $M_Q \geq 10^{12}~GeV$ the angular  
acceptance is $2 \pi$ (corresponding to SECS coming only from above), while for 
$M_Q \geq 10^{22}~GeV$ the geometrical acceptance is $4 \pi$.

\begin{figure}
\begin{center}
       \mbox{ \epsfysize=15cm
	\epsfbox{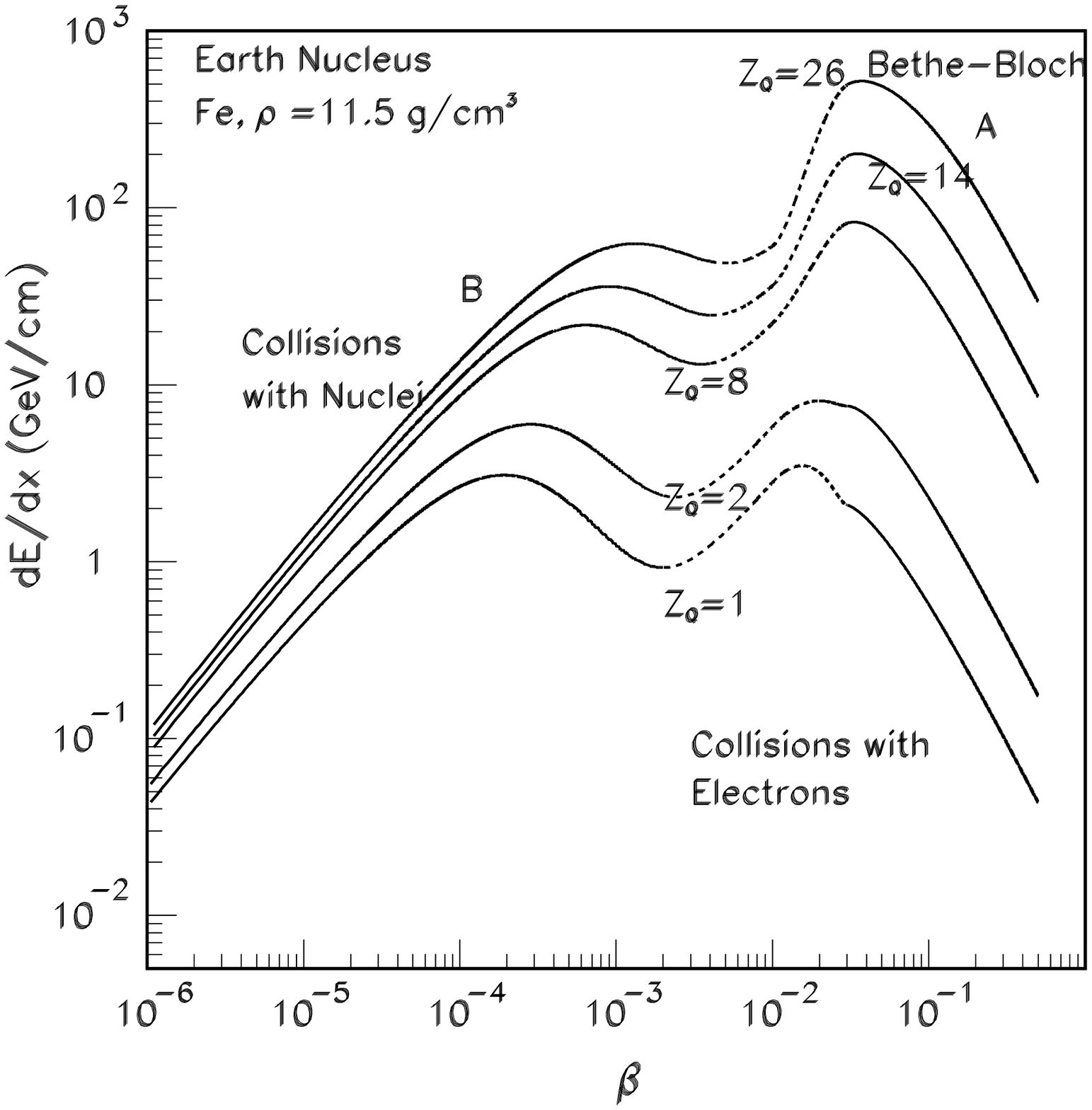}} 
\vspace{-1cm}
\caption{Energy losses of SECS versus $\beta$ in the Earth core. 
$Z_{Q}$ is the electric 
charge of the Q-ball. The possibility of emission of electromagnetic radiation 
at high $\beta \simeq 1$ was not considered.}
\label{fig:6}
\end{center}
\end{figure}

\begin{figure}
\begin{center}
        \mbox{ \epsfysize=15cm
             \epsfbox{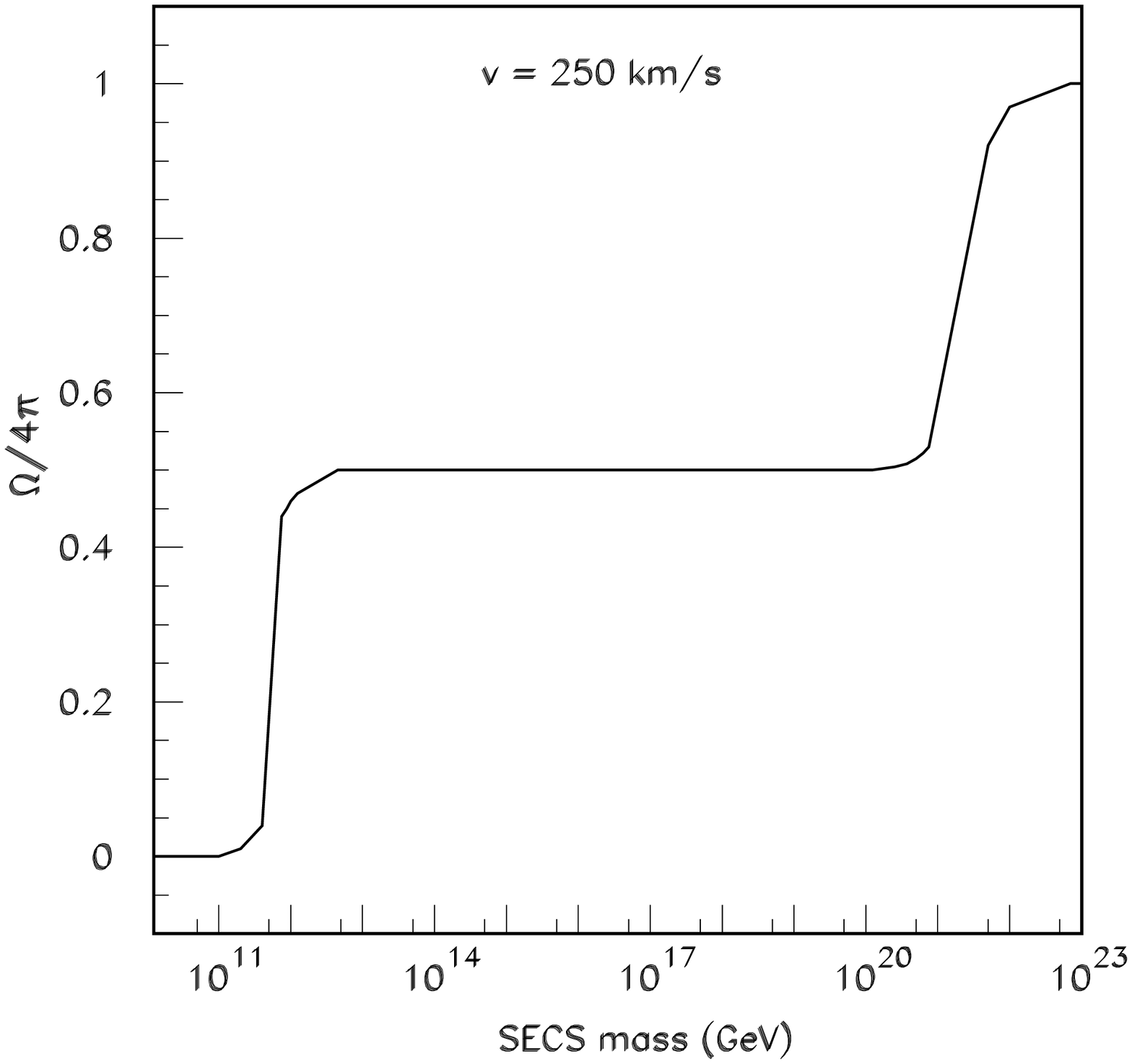}}
\vspace{-1cm}
\caption{Angular acceptance of the MACRO detector for Q-balls type 
SECS with electric charge $Z_{Q}=1$ and incoming velocity $v=250~km/s$,  
as function of the Q-ball mass $M_{Q}$. For $M_Q \geq 10^{12}~GeV$ the 
acceptance is $2\pi$ (only SECS from above), while for $M_Q \geq 10^{22}~GeV$ 
it is $4\pi$}
\label{fig:7}
\end{center}
\end{figure}

Fig. \ref{fig:6} shows the accessible regions  
in the plane ($M,~\beta$) for Q-balls type 
SECS  with $Z_{Q}=1$,  
coming from above (to the right of the dashed curve) and from below 
(to the right of the solid line), respectively. 
Notice that SECS 
coming from above (below) may reach MACRO only il they have masses larger than 
$\sim 10^{13}$ GeV ($\sim 10^{18}$ GeV).

\begin{figure}
\begin{center}
        \mbox{ \epsfysize=15cm
            \epsfbox{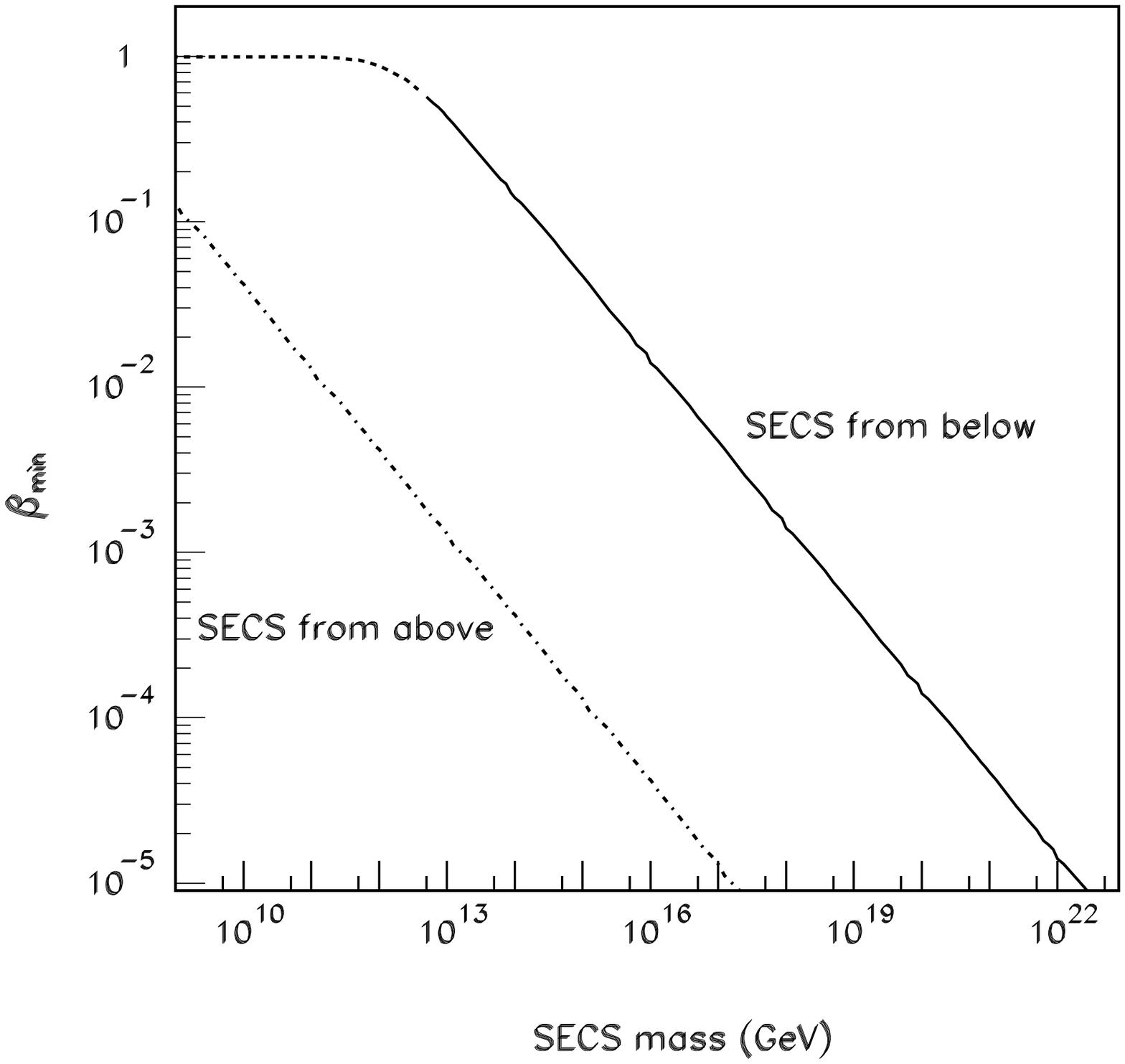}}
\vspace{-1cm}
\caption{Accessible region in the plane (mass, $\beta$) of
SECS with electric charge $Z_{Q}=1$. The solid (dotted-dashed) line, gives the 
minimum value of $\beta$ ($\beta_{min}$) that a SECS sould have to reach the 
MACRO detector from above (below); 
the accessible region to the upper right of the two curves. 
The dashed line is an extrapolation of the solid line to very large values 
of $\beta$.}
\label{fig:8}
\end{center}
\end{figure}

\subsection{Interactions of Q-balls type SENS}
The Q-ball interior of SENS is characterized by a 
large Vacuum Expectation Value  (VEV) of squarks,   
and may be of sleptons and Higgs fields \footnote{Also SECS have in their 
interior a large VEV of squarks, sleptons and Higgs fields; in addition they 
have a positive electric charge, and thus a coulomb potential which reduces 
the possibility of the catalysis of proton decay.} 
 \cite{Kusenko98A}. 
The $SU(3)_{c}$ symmetry is broken and deconfinement takes 
place inside the Q-ball.

If a nucleon enters the region of deconfinement, it dissociates into three 
quarks, some of  
which may  then become absorbed in the supersymmetric condensate [7-8]. 
The nucleon enters the $\tilde{q}$ condensate and giving  rise to  
processes like 
\begin{equation}
qq \rightarrow \tilde{q}\tilde{q}
\end{equation}

In practice one has

\begin{equation}
(Q) + Nucleon \rightarrow (Q+1) + pions \label{eq:13}
\end{equation}
and less probably, 
\begin{equation}
(Q) + Nucleon \rightarrow (Q+1) + kaons \label{eq:14}
\end{equation}

The pions carry the electric charge of absorbed nucleon 
(for example $\pi^{+}\pi^{0}$).

If it is assumed that the energy released in 
(\ref{eq:13}) and  (\ref{eq:14})   
is the same as in typical 
hadronic processes 
(about 1~GeV per nucleon), this energy is  carried by 2 or 3 pions (or   
kaons). 
The cross section 
for reactions (13) and (14) 
 are determined by the Q-ball radius  $R_{Q}$ \cite{Kusenko98A}
\begin{equation}
\sigma = \pi~R_{Q}^2 = \frac{16 \pi^2}{9}~M_{Q}^{-2}~Q^2 
\sim 6 \times 10^{-34}Q^{1/2} \left( \frac {1~TeV}{M_S} 
\right)^2~cm^2  \label{eq:15}
\end{equation}
The corresponding mean free path $\lambda$ is 
\begin{equation}
\lambda = \frac{1}{\sigma n}  \label{eq:16}
\end{equation}
where $n$ is the number of atoms per $cm^3$ in the traversed material. 
According to refs. [6-8] the 
energy loss of SENS moving with velocities in the range 
$10^{-4}<\beta <10^{-2}$ is constant and is given by 
\begin{equation}
\frac{dE}{dx} \sim \frac{\zeta}{\lambda} = \sigma n \zeta
\sim 6 \times 10^{-34}Q^{1/2}~n \zeta
\label{eq:17}
\end{equation}
where $\zeta = 1~({\rho}/1g~cm^{-3})~GeV$ is 
the energy released in one reaction.  
The energy loss of SENS is due to the energy released from the 
absorbed nucleon mass. Large mass SENS lose a small fraction of their 
kinetic energy and are able to traverse the earth without attenuation for all 
masses of our interest. Fig. \ref{fig:7} shows the energy loss of SENS with 
$10^{-4} \leq \beta \leq 10^{-2}$ plotted versus radius for $M_{S} = 1~TeV$.

Fig. \ref{fig:8} shows the mean free path $\lambda$ of Q-balls type SENS versus 
Q-ball number $Q$ for $M_{S}=1~TeV$.

\begin{figure}
\begin{center}
        \mbox{ \epsfysize=15cm
            \epsfbox{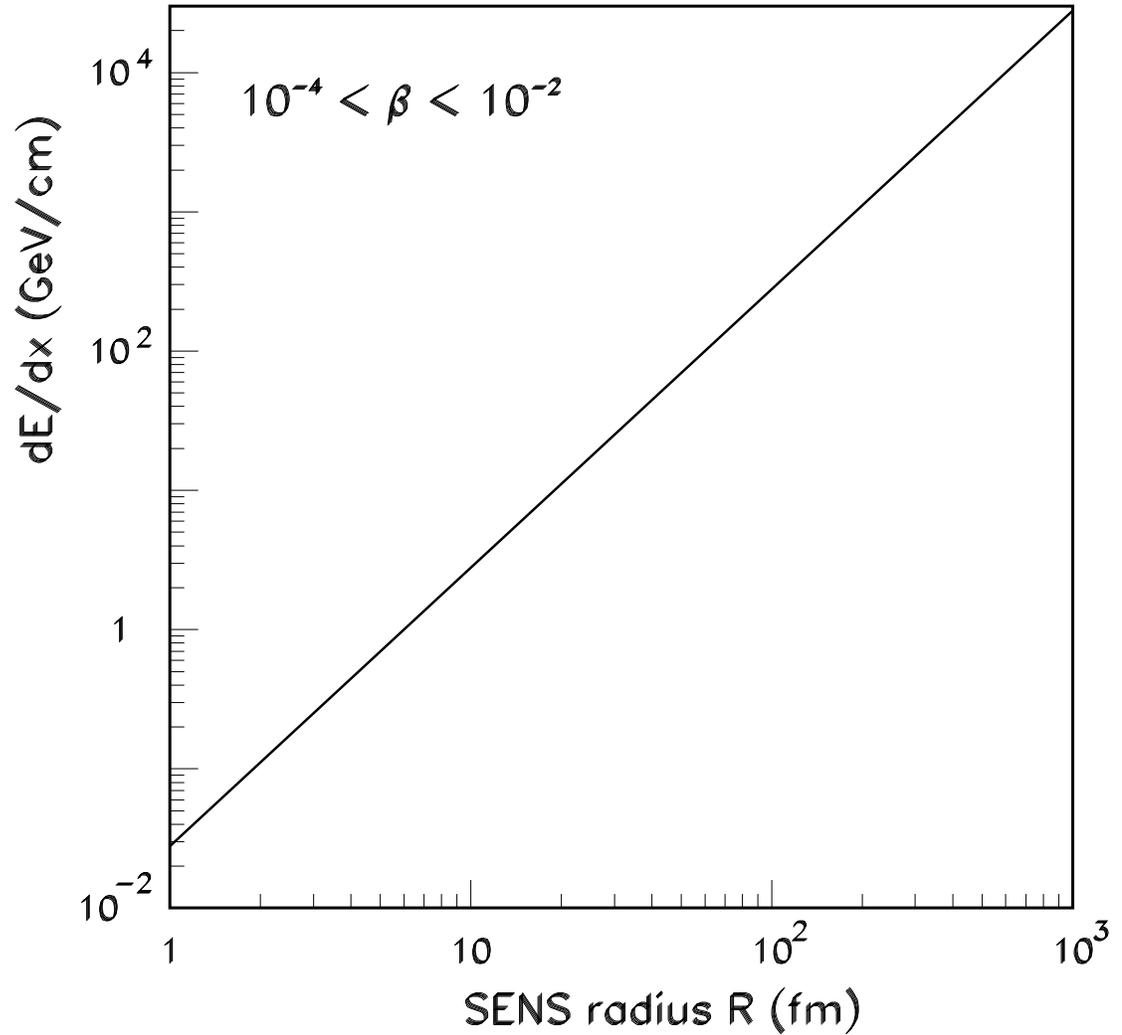}}
\vspace{-1cm}
\caption{Energy loss of Q-ball type SENS with $10^{-4} <\beta< 10^{-2}$ 
as function 
of the SENS radius $R$ in a material, Eq. \ref{eq:17}. This energy loss is 
in reality coming
 from the absorption of nucleons, see reactions (13),(14).}
\label{fig:9}
\end{center}
\end{figure}

\begin{figure}
\begin{center}
        \mbox{ \epsfysize=15cm
            \epsfbox{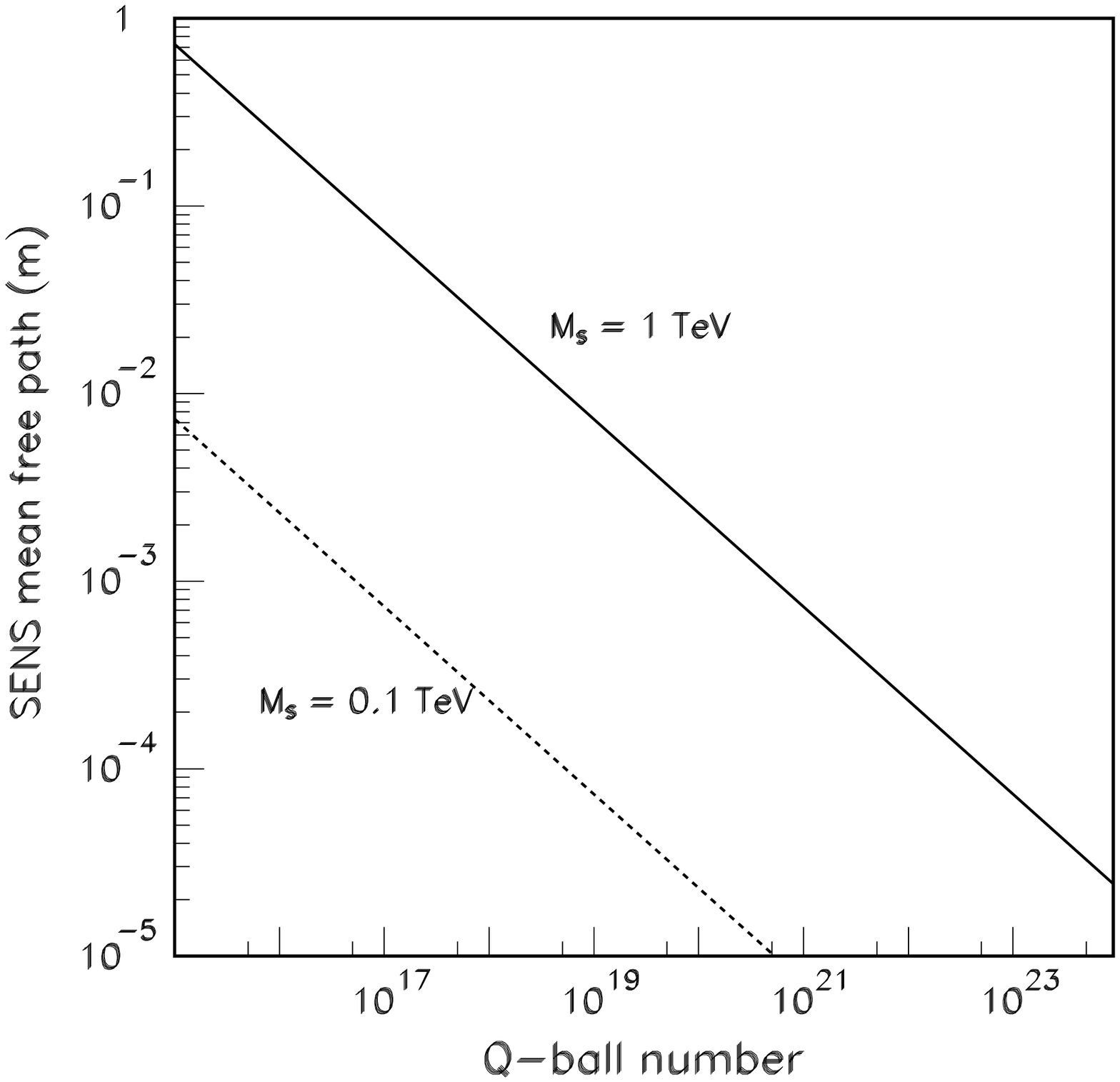}}
\vspace{-1cm}
\caption{Mean free path of Q-ball type SENS for $M_S=1~TeV$ and $M_S=0.1~TeV$
as function of 
the Q-ball number, Eq. \ref{eq:16}. The energy lost in each interaction is 
$\sim 1~GeV$ (absorption of one nucleon mass).}
\label{fig:10}
\end{center}
\end{figure}

\section{Sensitivity~to~SECS~of~scintillators,~streamer tubes and 
nuclear track detectors}
In this section we discuss the response of liquid scintillators, of streamer 
tubes
and of nuclear track detectors (in particular the MACRO CR39) to Q-balls, 
assuming that for $10^{-4} < \beta < 10^{-2}$ the 
electric charge of SECS is constant.

\subsection{Light yield of Q-balls type SECS}

The MACRO liquid sintillator has a density of 0.86~g/cm$^3$ and it is made of 
96.4 $\%$ of mineral oil, 3.6 $\%$ of pseudocumene, 1.44~g/l of PPO, 1.44~mg/l 
of bis-MSB and 40 mg/l of antioxydant [6,7].

For SECS we distinguish two contributions to the light yield in scintillators: 
the primary yight yield  and the secondary light yield.

{\bf \em The primary light yield:} it is due to the direct excitation (and 
ionization that occurs only for $\beta > 10^{-3}$)
produced by the SECS in the medium. The energy loss in the MACRO 
liquid scintillator is computed from the energy loss of protons in hydrogen 
and carbon \cite{D97}
\begin{equation}
\left(\frac{dE}{dx} \right)_{SECS}= \frac{1}{14} \left[~2 
\left(\frac{dE}{dx}\right)_{H} + 
12 \left(\frac{dE}{dx} \right)_{C} \right] = SP = \frac{SL \times SH}{SL + SH}
\label{eq:18}
\end{equation}
where $SP$ is the stopping power of SECS, which  reduces to the lowest stopping 
power (SL) at low 
$\beta$ and to the highest stopping power (SH) 
at high $\beta$.   The stopping power 
coincides with the Bethe-Bloch formula for electric 
energy losses at relatively high velocities. 

{\large \bf 1.} For electric charge $q=1e$ the energy loss of SECS in hydrogen and 
carbon is computed  
from \cite{Z77} adding an exponential factor coming from 
 the experimental data (see ref. [13]) on slow protons, see ref.
\cite{F87}.

{\bf i)}~For $10^{-5}  <   \beta < 5 \times 10^{-3}$ 
we have the following formula
\begin{equation}
\left(\frac{dE}{dx} \right)_{SECS}
= 1.3 \times 10^{5} \beta \;\; \left[~1- \exp \left( \frac{\beta}{
7 \times 10^{-4}} \right)^{2} \right]~~~\frac{MeV}{cm} \label{eq:19}
\end{equation}

{\bf ii)}~For $5 \times 10^{-3}  <   \beta < 10^{-2}$ we used the following 
formula  \cite{F87}
\begin{equation}
SP = SP_{H} + SP_{C} = \left(\frac{dE}{dx} \right)_{SECS} \label{eq:20}
\end{equation}
where
\begin{equation}
SP_{H} = \frac{SL_{H} \times SH_{H}}{SL_{H} + SH_{H}} \label{eq:21}
\end{equation}

\begin{equation}
SP_{C} = \frac{SL_{C} \times SH_{C}}{SL_{C} + SH_{C}} \label{eq:22}
\end{equation}
and
\begin{equation}
SL = A_{1}~E^{0.45},~~~~~~~~~SH = A_{2}~Ln \left( 1 + 
\frac{A_{3}}{E} + A_{4} E \right) \label{eq:23}
\end{equation}
where ($A_{i=1,4}$) are estimated in ref. [13]  and $E$ is 
the energy of a proton with velocity $\beta$.

{\large \bf 2.} For SECS with electric charge $q = Z_{1}e$ the energy losses for $ 10^{-5} < 
\beta < 10^{-2}$
are given by \cite{L61}
\begin{equation}
\left(\frac{dE}{dx} \right)_{SECS} = \frac{ 8 \pi e^{2} a_{0} \beta}
{\alpha} \frac{Z_{Q}^{7/6} n_{e}}
{(Z_{Q}^{2/3} + Z^{2/3})^{3/2}} \left[ ~1- \exp \left( - \frac{\beta}
{7 \times 10^{-4}} \right)^2 \right] \label{eq:24}
\end{equation}
where $Z$ is the atomic number of the target atom, $n_e$ the density 
of electrons  and $\alpha$ is the fine structure constant.

The primary light yield of SECS is given by \cite{D97}
\begin{equation}
\left(\frac{dL}{dx} \right)_{SECS} 
= A~ \left[ \frac{1}{1 + AB~\frac{dE}{dx}} \right]~\frac{dE}{dx} \label{eq:25}
\end{equation}
where $dE/dx$ is the total energy loss of SECS;   
$A$ is a conversion constant of the energy losses 
in photons (light yield) and $B$ is the 
parameter describing the saturation of the light yield; both  
parameters depend only on the velocity of SECS [12]. For example for the 
$\beta$-range, $5 \times 10^{-5} <\beta< 10^{-3}$, $A=0.067$ and 
$B=0.66$ cm/MeV.

{\bf \em The secondary light yield} arises from recoiling particles: 
we consider the elastic or quasi-elastic
recoil of hydrogen and carbon nuclei. The light yield $L_{p}$ from  
a hydrogen or carbon nucleus of
given initial energy $E$ is computed as

\begin{equation}
L_{p}(E)=\int^{E}_{0}\frac{dL}{dx}(\epsilon)S^{-1}_{tot} \, d\epsilon
\label{eq:26}
\end{equation}
where $S_{tot}$ is the sum of electronic and nuclear 
energy losses, $\epsilon$ is the excitation energy of the outer shell 
electrons.    
The  nuclear energy losses are given in ref. \cite{W77}.
The secondary light
yield is then
\begin{equation}
\left(\frac{dL}{dx}\right)_{\mbox{secondary}}=N\int^{T_{m}}_{0}L_{p}(T)
\frac{d\sigma}{dT} \, dT \label{eq:27}
\end{equation}
where $N$ is the number density of nuclei in the medium  
$T_{m}$ is the maximum energy transferred and
 {\Large $\frac{d\sigma}{dT}$} is the
differential scattering cross section, given in ref. \cite{L77}.

Fig. \ref{fig:9} shows the light yield of 
SECS in the MACRO liquid scintillator; for 
reference are also indicated the light yields of MMs with $g =g_D$ and fast 
muons.

\begin{figure}
\begin{center}
	\mbox{ \epsfysize=18cm
	\epsfbox{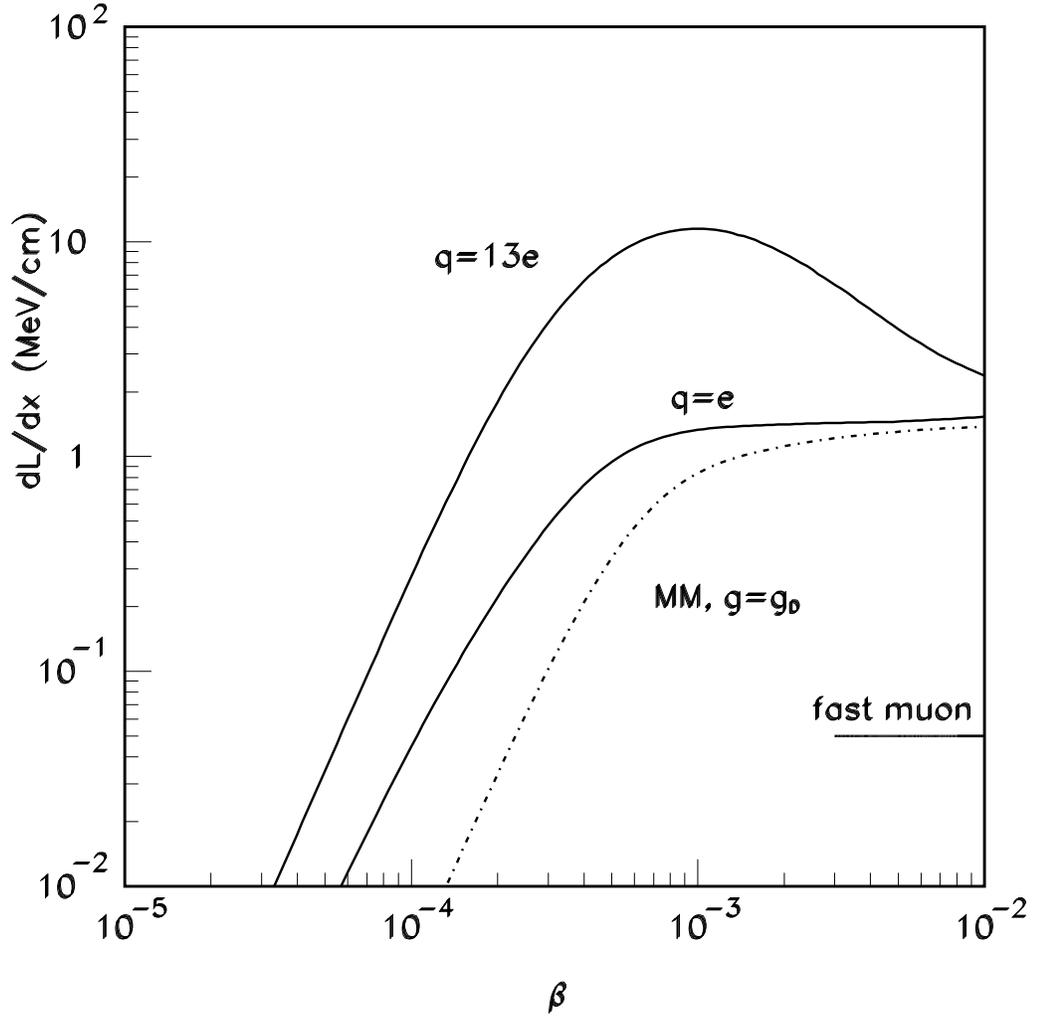}}
\vspace{-4cm}
\caption{Light yield of SECS (solid lines)
in the MACRO liquid scintillator as function of 
the SECS velocity;  $q$ is the net positive electric charge of the 
SECS. For comparison we give also the light yield of MMs with 
$g=g_D$ (dotted-dashed line) and of fast muons. 
For $2 \times 10^{-4}<\beta<3 \times 10^{-3}$ there is no saturation of the 
light yield for $q=13e$ because it is the recoil nucleus which contributes 
to the light yield.}
\label{fig:11}
\end{center}
\end{figure}

\subsection{Energy losses of SECS in streamer tubes}
The composition of the gas in the  MACRO limited streamer tubes is 73\% 
helium and   27\% n-pentane, in volume \cite{Ouchrif98}. 
The pressure is about one atmosphere and the resulting density
is low (in comparison with the density of the
other detectors): $ \rho_{gas}=0.856~\mbox{mg/cm}^{3}$. 
 The energy losses of MMs in the streamer tubes have been discussed in ref.  
[6,7].

The ionization energy losses of SECS in the MACRO streamer tubes for
$10^{-3} < \beta < 10^{-2}$ are computed with the same general 
procedure used 
for scintillators, using the density and the chemical composition of 
streamer tubes. 

For $q =13e$ the ionization 
energy losses are calculated as in ref. \cite{L61}
(we have omitted the exponential factor which takes into account the 
energy gap in organic scintillators).

The threshold for ionizing n-pentane occurs for  $\beta \geq 2 \times 10^{-3}$. 

Fig. \ref{fig:10} shows the ionization energy losses of SECS with electric 
charges $q=1$ and $q=13$ plotted versus $\beta $.

\begin{figure}
\begin{center}
	\mbox{ \epsfysize=15cm
		\epsfbox{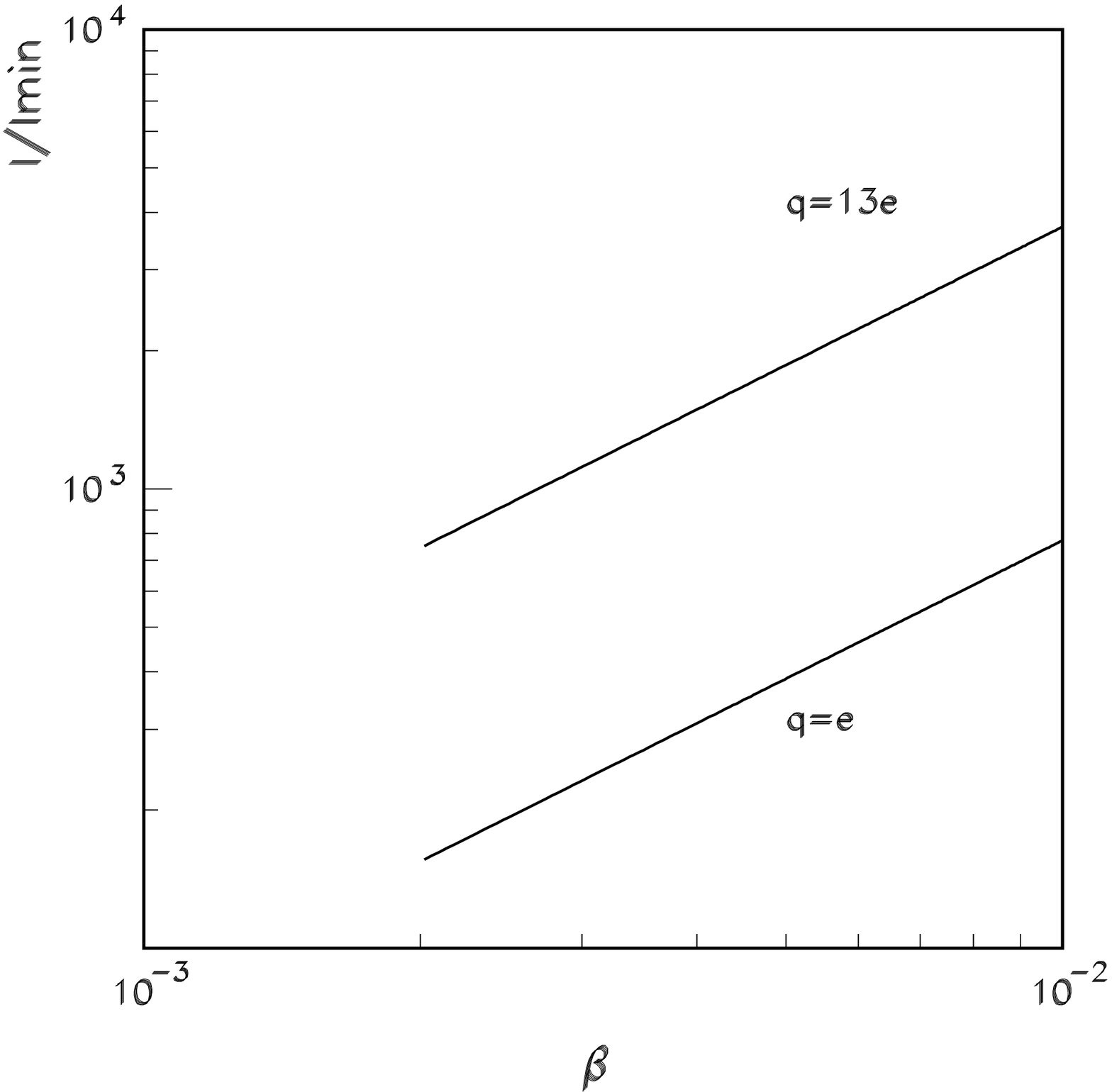}}
\vspace{-1cm}
\caption{Ionization energy losses by SECS versus $\beta$
in the limited streamer tubes, 
filled with 
$73 \%$  He and $27 \%$  n-pentane, and  with density 
$\rho_{gas} = 0.856~ mg/cm^3$,  
relative to the ionization produced by a minimum ionizing particle, 
$I_{min} = 2.2~MeV cm^2 g^{-1}$}
\label{fig:12}
\end{center}
\end{figure}

\subsection{Restricted energy losses of SECS in the 
nuclear track detector CR39}
The quantity of interest for nuclear track detectors is the
Restricted Energy Loss (REL), that is, the energy deposited within $\sim$~
100~\AA~~diameter from the track.

The REL in CR39 has already been computed for MMs of $g=g_{D}$ and
$g=3g_{D}$ and for dyons with 
$q = e$, $g=g_{D}$ in ref. \cite{R83}. We have
checked these calculations and extended them to other cases of interest, see 
ref.  \cite{Ouchrif98}.

The chemical composition of CR39 is
$(\mbox{C}_{12}\mbox{H}_{18}\mbox{O}_{7})_n$, and the
density is 1.31~$\mbox{g/cm}^3$.
For the computation of the REL, only energy transfers to atoms 
larger than $12$~eV are
considered, because it is estimated that $12$~eV are necessary to break the
molecular bonds \cite{D97}.

At {\em low velocities} ($3 \times 10^{-5}<\beta<10^{-2}$) there are
two contributions to REL: the ionization and the atomic recoil 
contributions.

{\em The ionization contribution,} which becomes important only for $\beta > 
2 \times 10^{-3}$,   
was computed
with Ziegler's fit to the experimental data \cite{Z77}. 

{\em The atomic recoil contribution,}  which is important to REL for 
low $\beta$ values was calculated  
using the  interaction potential between an atom and a SECS \cite{W77} 

\begin{equation}
V(r) = \frac{Z_{Q} Z  e^{2}}{r} \phi (r) \label{eq:28}
\end{equation}
where $r$ is the distance between the core of SECS and the target atom, 
$Z_{Q}e$ is 
the electric charge of the SECS core,  
$Z$ is the atomic number of the target atom. The 
function $\phi (r)$ is the screening function given by \cite{W77} 
\begin{equation}
\phi (r) = \sum_{1}^{3} C_{i}~exp[- \frac{b_{i} r}{a}] \label{eq:29}
\end{equation}
where $b_{i}$ and $c_{i}$ are semiempirical constants given in Table 1 
of ref. [16] and 
the coefficients $C_i$  are restricted such that 
$\sum_{1}^{3} C_{i} =1  \label{eq:31}$, 
$a$ is the screening length of Eq. [12].

Assuming the validity of the potential of Eq. 29, we calculate the 
relation between the scattering angle $\theta$ and the impact parameter b. 
From this
relation, the differential cross section $\sigma (\theta )$ is obtained
as \cite{D97}
\begin{equation}
\sigma (\theta ) = -(db/d\theta)\cdot b/\sin \theta \label{eq:32}
\end{equation}
\hspace{0.6cm}
The relation between the transferred kinetic energy $K$ and the scattering 
angle $\theta$
is given by the relation
\begin{equation}
K = 4E_{inc} \sin^{2} (\theta/2) \label{eq:33}
\end{equation}
where $E_{inc}$ is the energy of the atom relative to the SECS 
in the center of mass system. The restricted 
energy losses are finally obtained by
integrating the transferred energies as
\begin{equation}
-\frac{dE}{dx}  = N \int \sigma(K) \, dK \label{eq:34}
\end{equation}
where $N$ is the number density of atoms in the medium, $\sigma(K)$ is the
differential cross section as function of the transferred kinetic energy $K$.
 
Fig. \ref{fig:11} shows the computed REL in CR39 for SECS with $q=1e$ and 
$q=13e$ 
plotted versus $\beta$.

\section{Conclusions}
Supersymmetric generalizations of the Standard Model allow for 
stable non-topological solitons, called  Q-balls, which may be considered 
as bags of 
squarks and sleptons and thus have non-zero 
baryon and lepton numbers; they have positive electric charge (SECS) or 
neutral (SENS) small masses [1-3]. 
The Q-ball could be  
produced in the Early Universe, could affect the nucleosynthesis 
of light elements, and could 
lead to a variety of other astrophysical and cosmological consequences.

In this paper, we computed the energy losses of Q-balls of type SENS and 
SECS. Using these energy losses and a rough model of the Earth's  composition 
and density profiles, we have computed the angular acceptance of 
the an underground detector at the Gran Sasso Laboratory 
for Q-balls type SECS with $v=250~km/s$ as  function of 
the Q-ball mass $M_{Q}$. We have calculated the accessible region in the plane  
(mass,  velocity) of SECS reaching the MACRO detector from above and from below.

We also presented an analysis of the energy deposited in the MACRO 
subdetectors: 
scintillators, streamer tubes and CR39 nuclear track detectors by SECS, in 
forms useful for their detection. In particular we computed the light yield 
in scintillators, the ionization in streamer tubes and the 
REL in nuclear track detectors. 

The three MACRO subdetectors are sensitive to SECS with $\beta \sim 10^{-3}$ 
and masses larger than $10^{13}~GeV$ ($10^{19}~GeV$) for SECS coming 
from above (below).  
A flux upper limit may  
be obtained from MACRO 
at the level of few times $10^{-16}~~cm^{-2}s^{-1}sr^{-1}$.

SENS are more difficult to detect. MACRO scintillators could detect them, the 
streamer tubes have a limited efficiency and the CR39 detectors 
cannot see them. 
\par
\vspace{1cm}
\newpage
{\bf Acknowledgements:} 
We gratefully acknowledge the cooperation of many members of the MACRO 
collaboration, in particular of all the members of Bologna group. We  
thank A. Kusenko for stimulating discussions.

\begin{figure}
\begin{center}
        \mbox{ \epsfysize=15cm
        \epsfbox{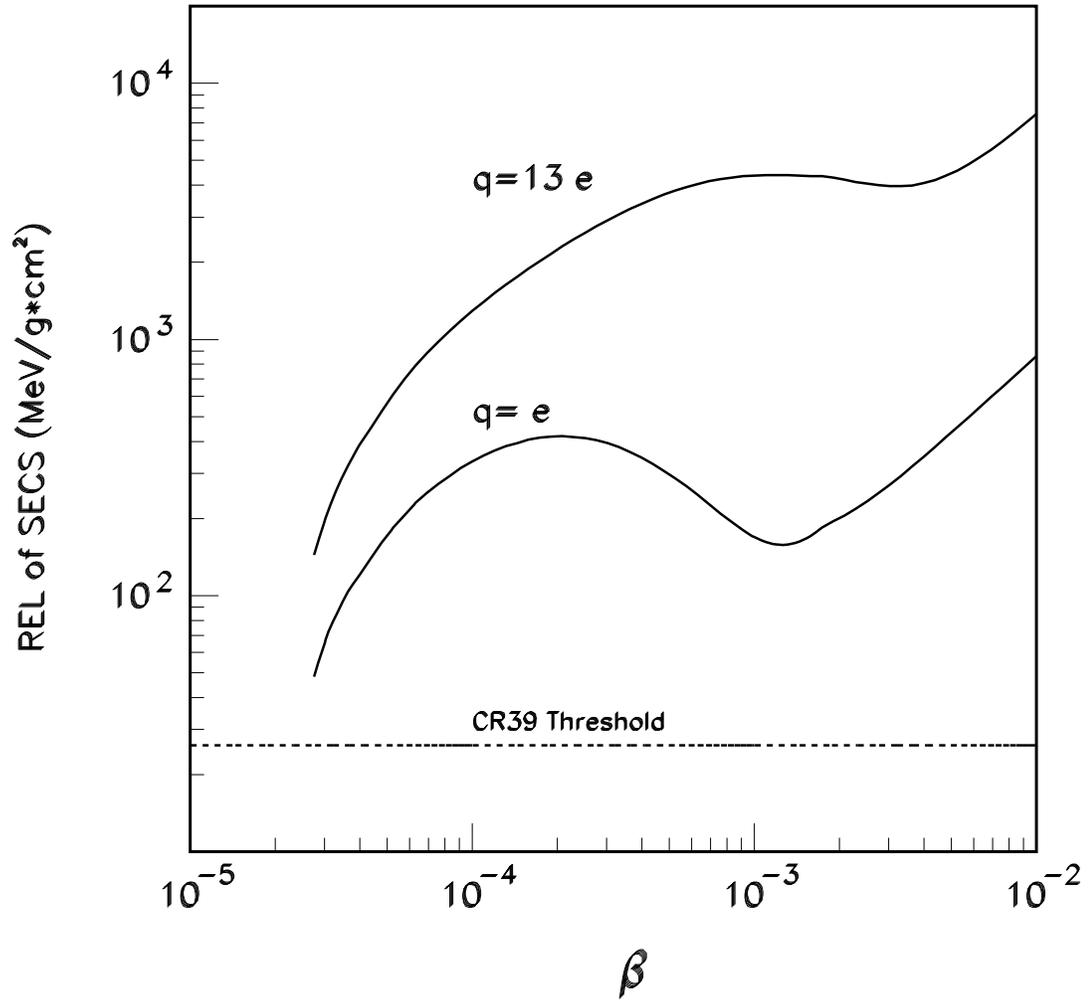}}
\vspace{-1cm}
\caption{Restricted Energy Losses of SECS with charges $e$ and $13e$ 
as function of their velocity in the 
nuclear track detector CR39. The detection threshold for the MACRO CR39 is 
also shown, ref.~[7].}
\label{fig:13}
\end{center}
\end{figure}

\end{document}